\documentclass[12pt]{article}
\usepackage{amssymb,graphicx,epsfig,psfig}
\begin{document}

\baselineskip=.22in
\renewcommand{\baselinestretch}{1.2}
\renewcommand{\theequation}{\thesection.\arabic{equation}}

\begin{flushright}
{\tt hep-th/0508027}
\end{flushright}

\vspace{5mm}

\begin{center}
{{\Large \bf Global and Local D-vortices}\\[12mm]
Yoonbai Kim\\[2mm]
{\it BK21 Physics Research Division and Institute of Basic Science,}\\
{\it Sungkyunkwan University, Suwon 440-746, Korea}\\
{\tt yoonbai@skku.edu}\\[5mm]
Bumseok Kyae\\[2mm]
{\it School of Physics, Korea Institute for Advanced Study,\\
207-43, Cheongryangri-Dong, Dongdaemun-Gu, Seoul 130-012, Korea}\\
{\tt bkyae@kias.re.kr}\\[5mm]
Jungjai Lee\\[2mm]
{\it Department of Physics, Daejin University,
GyeongGi, Pocheon 487-711, Korea}\\
{\tt jjlee@daejin.ac.kr}
}
\end{center}

\vspace{5mm}

\begin{abstract}
Codimension-two objects on a system of brane-antibrane are studied
in the context of Born-Infeld type effective field
theory with a complex tachyon and U(1)$\times$U(1) gauge fields.
When the radial electric field is turned on in D2${\bar {\rm D}}$2,
we find static regular global and local D-vortex solutions
which could be candidates of straight cosmic D-strings
in a superstring theory. A natural extension to DF-strings is briefly
discussed.
\end{abstract}


\newpage

\setcounter{equation}{0}
\section{Introduction and Model}

Cosmic strings arise as vortices in field theories, mainly when
either global or gauge U(1) symmetry is spontaneously broken.
The point-like vortices in 2-dimensional space become long straight strings
in 3-dimensional space simply by stretching them along the extra-spatial
direction with maintaining the 2-dimensional field configurations
independent of the straight third spatial coordinate.
Thus, if these straight strings are gravitating, it is almost the same as
gravitating the point-like vortices in planar gravity, which results in
conic geometry at asymptote. These static properties of cosmic strings
constitute a main basis for studying formation and evolution of
straight and wiggly cosmic strings
and their networks, which could be left as cosmological fossils at the present
Universe~\cite{VS,Kibble:2004hq}.

Since almost all of the ideas from vortices in local gauge
theories with Higgs mechanism can carry over to superstrings, it
seems natural to consider macroscopic superstring as a candidate
of cosmic strings. This possibility was firstly taken into account
in Ref.~\cite{Witten:1985fp}. However, it was excluded by the
following reasons as production of large inhomogeneity in the
cosmic microwave background, dilution of string density through
inflation, and instability to keep long strings. Development of
D-branes during the last decade encourages us to tackle again this
issue with cosmic D- and
DF-strings~\cite{Dvali:2003zj,Copeland:2003bj,Polchinski:2004ia}
and the related
topics~\cite{Halyo:2003uu,Leblond:2004uc,Jackson:2004zg}.

For the description of D- and DF-strings, an efficient approach is
to employ the method of effective field theory from string theory.
As stated in the first paragraph, the first step for constructing
cosmic D- and DF-strings in the context of effective field theory
is to find point-like vortex solutions in the plane, particularly
static and regular
solutions~\cite{Jones:2002cv,Dvali:2003zh,Sarangi:2002yt}. In the
scheme of effective field theory, instability of brane-antibrane
system is represented by a complex tachyon field $(T,\bar{T})$,
and this tachyon is indispensable for the generation of
topological defects. Since this system possesses U(1)$\times$U(1)
gauge symmetry, we need two gauge fields, $A^{a}_{\mu},\;a=1,2$.
Separation of the brane and the antibrane is described by scalar
fields $X^{I}_{a}$ corresponding to the transverse coordinates of
individual branes. Among various proposed tachyon effective
actions~\cite{Kraus:2000nj,Jones:2002si}, we shall deal with
Dirac-Born-Infeld (DBI) type effective action of the D$p{\bar {\rm
D}}p$ system in Ref.~\cite{Sen:2003tm},
\begin{eqnarray}
S&=&-{\cal T}_{p}\int d^{p+1}x\, V(T,X^{I}_{1}-X^{I}_{2}) \nonumber\\
&& \hspace{15mm}\times
\left\{\,\sqrt{-\det \left[g_{\mu\nu}+F_{\mu\nu}^{1}+\partial_{\mu} X^{I}_{1}
\partial_{\nu}X^{I}_{1}+({\overline {D_{\mu}T}}D_{\nu}T
+{\overline {D_{\nu}T}}D_{\mu}T)/2
\right]}\right.
\nonumber\\
&&\hspace{20mm}\left.
+\sqrt{-\det \left[g_{\mu\nu}+F_{\mu\nu}^{2}+\partial_{\mu} X^{I}_{2}
\partial_{\nu}X^{I}_{2}+({\overline {D_{\mu}T}}D_{\nu}T
+{\overline {D_{\nu}T}}D_{\mu}T)/2
\right]}
\,\,\right\},
\label{dea}
\end{eqnarray}
where ${\cal T}_{p}$ is the tension of the D$p$-brane,
$F_{\mu\nu}^{a}=\partial_{\mu}A^{a}_{\nu}-\partial_{\nu}A^{a}_{\mu}$,
and $D_{\mu}T=(\partial_{\mu}-iA_{\mu}^{1}+iA^{2}_{\mu})T$.
For small tachyon amplitude $\tau$ from $T=\tau e^{i\chi}$,
behavior of the tachyon potential is
\begin{equation}\label{ptx}
V(T,X^{I}_{1}-X^{I}_{2})=1-
\left[\frac{1}{R^{2}}-\sum_{I}(X^{I}_{1}-X^{I}_{2})^{2}
\right]\tau^{2}+{\cal O}(\tau^{4}),
\end{equation}
where $R$ is $\sqrt{2}$ in superstring theory.

In the context of DBI-type effective action describing D-brane
systems with instability, there have been much study on
codimension-one solitons (codimension-one branes), particularly
tachyon
kinks~\cite{Lambert:2003zr,Sen:2003tm,Kim:2003in,Brax:2003rs,Kim:2003ma}.
For vortices (codimension-two branes), only the singular local
vortex solution with finite energy was constructed from the DBI
action (\ref{dea})~\cite{Sen:2003tm}, and regular tachyon vortex
solutions were obtained in local field theory action with
quadratic derivative terms and polynomial tachyon
potential~\cite{Hashimoto:2002xe,Jones:2002cv,Jones:2002si,Dvali:2003zj}.

It is well-known that fundamental strings attached to a D-brane
becomes flux tube solutions (spikes or BIons) with nonzero
thickness in nonlocal field theory of
DBI-action~\cite{Callan:1996dv}. Thickness of a tachyon kink
becomes nonzero when the constant DBI electric field transverse to
the codimension-one D-brane is tuned on, and, it becomes, at its
critical value, thick static regular BPS tachyon
kink~\cite{Kim:2003in,Kim:2003ma}. In this paper, we show that
both global and local D-vortices have zero-thickness without DBI
electric field, but global and local D-vortices with electric flux
become regular with nonzero thickness. These regular
configurations are interpreted as a nonzero size D0 on D2${\bar
{\rm D}}$2 on which fundamental strings are attached.

In section 2, without a U(1) gauge field, we discuss static singular
global D-vortex configuration with zero radial electric field and static regular
global D-vortex with nonzero radial electric field.
In section 3, with a U(1) gauge field, we find the singular and regular
local D-vortex solutions.
Section 4 involves discussions on RR-charge and extension of D-vortices to
straight stringy objects in D3${\bar {\rm D}}$3 system.
We conclude in section 5 with summary of the obtained results and
discussions for further study.

\setcounter{equation}{0}
\section{Global D-Vortices}
Let us consider D$p{\bar {\rm D}}p$ system in the coincidence limit
of two branes, $X^{I}_{1}=X^{I}_{2}$,
with fundamental strings.
Then the macroscopic fundamental strings in fluid state are represented by
electric fluxes along their directions.
Vortex-like codimension-two objects of our interest
could be interpreted as D$(p-2)$-branes. For description of the global
vortex-like objects,
the gauge fields and their field strengths should behave as
$A_{\mu}^{1}=A_{\mu}^{2}=A_{\mu}$ and $F_{\mu\nu}^{1}=F_{\mu\nu}^{2}
=F_{\mu\nu}$ in the action (\ref{dea}).
Then the action (\ref{dea}) in (1+$p$)-dimensions becomes
\begin{equation}\label{acg}
S=-2{\cal T}_{p}\int d^{p+1}x\, V(\tau)\sqrt{-\det X} \, ,
\end{equation}
where
\begin{equation}\label{Xm}
X_{\mu\nu}=g_{\mu\nu}+F_{\mu\nu}
+(\partial_{\mu}{\bar T}\partial_{\nu}T+\partial_{\nu}{\bar T}\partial_{\mu}T)
/2 .
\end{equation}

Additionally we assume that the produced D$(p-2)$-branes are flat
and all the transverse degrees are frozen.
Then we can neglect dependence of the transverse coordinates and
it is enough to find D0-branes from flat D$2{\bar {\rm D}}2$.
In the context of solitons in the effective theory, it is
translated as point-like vortices on a plane. We will call this vortex
as {\it D-vortex} in what follows.

When dependence of the transverse scalar fields $X^{I}_{a}$ disappears
in (\ref{ptx}), the tachyon potential $V$ is invariant under the
exchange symmetry of the brane and antibrane and coincides with
that of an unstable D$p$-brane~\cite{Sen:1999mg}.
Since it measures varying tension of the
${\rm D2}\bar{{\rm D}}2$ system and then universality allows any
runaway tachyon potential connecting the following
two boundary values smoothly and monotonically~\cite{Sen:1999xm}
\begin{equation}\label{bdpo}
V(\tau=0)=1\quad {\rm and} \quad V(\tau=\infty)=0
\end{equation}
which has a ring of degenerate minima at infinite tachyon amplitude.
We will use a specific
potential~\cite{Kutasov:2003er,Buchel:2002tj}
for the analysis for vortex solutions
\begin{equation}\label{V3}
V(\tau)=\frac{1}{{\rm cosh}\left(\frac{\tau}{R}\right)},
\end{equation}
but the obtained results are the same under the assumption of any symmetric
tachyon potential, $V(\tau)=V(-\tau)$,
satisfying (\ref{bdpo}) with $V(\tau)\sim e^{-\tau/R}$ for large
$\tau$~\cite{Sen:2002an}. Note that soliton solutions have been firstly
investigated under the action with quadratic derivative kinetic term and
runaway potential in the context of quintessence~\cite{Cho:1998jk}.

\subsection{Singular global D-vortex}

The simplest D-vortex configuration is that of static $n$ vortices
superimposed at a point.
It is convenient to use radial coordinates $(t,r,\theta)$ with
$g_{\mu\nu}={\rm diag}(-1,1,r^2)$ and to choose the vortex point as
the origin $r=0$.
Ansatz for the $n$ superimposed vortices is
\begin{equation}\label{ans}
T(r,\theta)=\tau(r)e^{in\theta},
\end{equation}
and that for D-vortices require vanishing $F_{\mu\nu}$.
Substituting these into the effective action (\ref{acg}), we have
\begin{equation}\label{actr}
\frac{S}{\int d^{p-2}x_{\perp}}
= -2{\cal T}_{p}\int dtdrd\theta r V(\tau)
\sqrt{\left(1+\tau^{'2}\right)\left(1+\frac{n^2}{r^2}\tau^2\right)}\, ,
\end{equation}
where $'$ denotes $d/dr$.

When we have a topological D-vortex solution
of vorticity $n$, the tachyon amplitude is given by a monotonic increasing
function connecting the following boundary values smoothly.
The boundary condition at the origin is forced by the ansatz (\ref{ans})
\begin{equation}\label{bd0}
\tau(r=0)=0,\qquad (n\ne 0),
\end{equation}
and that at infinity is read from runaway nature of the tachyon potential
(\ref{bdpo})--(\ref{V3})
\begin{equation}\label{bdin}
\tau(r\rightarrow \infty)\rightarrow \infty.
\end{equation}

First of all, we verify nonexistence of regular global D-vortex solution
satisfying the boundary conditions (\ref{bd0})--(\ref{bdin}) by examining
the tachyon equation from the action (\ref{actr}):
\begin{equation}\label{teqr}
\frac{1}{r}\frac{d}{dr}\left[
\frac{rV}{\sqrt{X}}\left(1+\frac{\tau^2}{r^2}n^2\right)~\tau'\right]
-\frac{V}{\sqrt{X}}(1+\tau^{'2})\tau\frac{n^2}{r^{2}}
=\sqrt{X}~\frac{dV}{d\tau} ,
\end{equation}
where
\begin{equation}
X=\left(1+\tau^{'2}\right)\left(1+\frac{n^2 \tau^2}{r^2}\right).
\end{equation}
Physical properties are read from non-vanishing components of the
energy-momentum tensor
\begin{eqnarray}
T^{t}_{\; t}&=&-\frac{2{\cal T}_{p}V}{\sqrt{X}}
\left[\left(1+\tau'^{2}\right)\left(1+\frac{n^2}{r^2}\tau^2\right) \right],
\label{fttt}\\
T^{r}_{\; r}&=&-\frac{2{\cal T}_{p}V}{\sqrt{X}}
\left(1+\frac{n^{2}}{r^{2}}\tau^{2}\right),
\label{ftrr}\\
T^{\theta}_{\; \theta}&=&-\frac{2{\cal T}_{p}V}{\sqrt{X}}
\left(1+\tau'^{2}\right).
\label{fttht}
\end{eqnarray}

Suppose that there exists single topological tachyon vortex
solution of vorticity $n$ ($\neq 0$), satisfying the boundary
conditions (\ref{bd0})--(\ref{bdin}). We try a power-series and a
logarithmic solution for sufficiently large $r$, assuming each
leading behavior of the tachyon amplitude is
\begin{equation}\label{tpin}
\tau(r)\approx \left\{
\begin{array}{ll}
\displaystyle{ \tau_{\infty}r^{k}}
& \qquad (k\ge 0), \\
\displaystyle{ \tau_{\infty}{\rm ln}r } &.
\end{array}
\right.
\end{equation}
Inserting it into the tachyon equation (\ref{teqr}), we have an approximate
equation from the leading terms
\begin{equation}\label{con}
\left\{
\begin{array}{ll}
\displaystyle{ C_{k}r^{-k} \approx
-\frac{1}{R}}
& C_{k}=0~\mbox{for}~k> 1, \\
\displaystyle{\frac{(1-n^{2})\tau_{\infty}}{1+n^{2}\tau_{\infty}^{2}}
\frac{1}{r} \approx -\frac{1}{R} }
& \mbox{for}~k=1~\mbox{and}~n\ne 1,\\
\displaystyle{-\frac{\tau_{\infty}^2}{R}\approx
-\frac{1+\tau_{\infty}^{2}}{R}}
& \mbox{for}~k=1~\mbox{and}~n=1,\\
\displaystyle{ \tau_{\infty}(k^{2}-n^{2})r^{p-2}\approx
-\frac{1}{R}}
& \mbox{for}~0<k<1,\\
\displaystyle{ -n^{2}\tau_{\infty}\frac{\ln r}{r^{2}}\approx
-\frac{1}{R}} & \mbox{for}~\tau(r)\approx \tau_{\infty}\ln r  .
\end{array}
\right.
\end{equation}
All five cases in the left-hand and right-hand sides of
(\ref{con}) lead to contradiction. An exponentially growing
$\tau(r)$ for sufficiently large $r$ also turns out to arrive at
contradiction. Thus, there does not exist static regular single
global topological vortex connecting $\tau(r=0)=0$ and
$\tau(r=\infty)=\infty$ monotonically. Note that the proof of
nonexistence holds for all the tachyon potentials only if
$V(\tau)\sim e^{-\tau/R}$ for large $\tau$.

This phenomenon of nonexistence seems familiar since it also appears
in tachyon kinks and tachyon tubes where the form of static regular
solutions are given by
array of kink-antikink or array of
tube-antitube~\cite{Lambert:2003zr,Kim:2003in,Brax:2003rs,Kim:2003uc}
instead of single kink or single tube.
Despite of the nonexistence of regular single tachyon kink, the constructed
singular single tachyon kink configuration~\cite{Sen:2003tm} was reconciled as
zero thickness limit of a single BPS kink part of the
array~\cite{Kim:2003in,Brax:2003rs,Kim:2004xk}.

Following the wisdom of boundary string field theory of
D2${\bar {\rm D}}$2-system~\cite{Kraus:2000nj}, let us try a linear tachyon
configuration
with infinite slope, $\tau(r)=r/\epsilon ,~~\epsilon\rightarrow 0$.
Then pressure components (\ref{ftrr})--(\ref{fttht})
have the same value even at the origin,
$T^{r}_{\;r}(0)=T^{\theta}_{\;\theta}(0)=-2{\cal T}_{2}$
and $T^{r}_{\;r}(r\ne 0)=T^{\theta}_{\;\theta}(r\ne 0)=0$,
and the conservation law of energy-momentum tensor equivalent to
the tachyon equation (\ref{teqr}) becomes simple in the limit of
$\epsilon\rightarrow 0$:
$T^{r'}_{\; r}+T^{r}_{\; r}/r
-T^{\theta}_{\;\theta}/r=T^{r'}_{\; r}=0$.
This nonvanishing pressure implies that
the obtained singular D-vortex may not be a BPS object
and it is different from
the cases of thin kink~\cite{Lambert:2003zr,Kim:2003in} or
thin tube~\cite{Kim:2004xk}, where transverse pressure vanishes $T_{xx}=0$.
Energy density $-T^{t}_{\;t}$ (\ref{fttt}) is peaked at the origin
like a $\delta$-function and the corresponding tension of the thin D0-brane
is given the following decent relation
\begin{eqnarray}
2{\cal T}_{0}^{g}&\equiv &\int_{0}^{2\pi}d\theta\int_{0}^{\infty}dr\, r
(-T^{t}_{\; t})\\
&\stackrel{\tau=r/\epsilon}{\approx}&2(2\sqrt{\pi K}R)^{2}{\cal T}_{2}.
\label{cata}
\end{eqnarray}
The Catalan's constant $K$ is almost unity, $K\equiv
\int_{0}^{\infty} dy y/2{\rm cosh}y\approx 0.91596...$, and so it implies
that the decent relation (\ref{cata}) is not exact even in the limit
of zero thickness, i.e., $2\sqrt{\pi K}/\pi=1.0799...$ means about 8$\%$
error.

\subsection{Regular global D-vortex with electric flux}

\begin{figure}[ht]
\centerline{\epsfig{figure=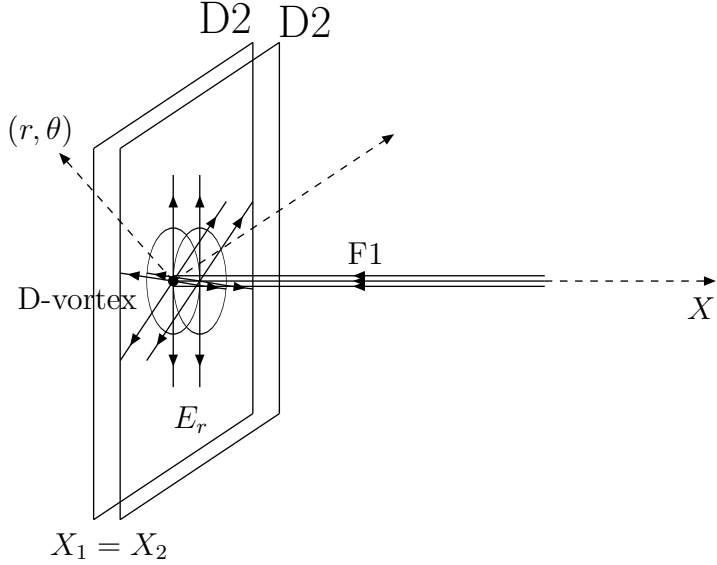}} \caption{D2 and ${\bar
{\rm D}}2$ with fundamental strings represented by $F_{0r}=E_r(r)$
in the coincidence limit of $X^{I}_{1}=X^{I}_{2}$. Both the
D-vortex and fundamental string are superimposed at the origin
($r=0$).} \label{fig1}
\end{figure}

Let us take into account D2${\bar {\rm D}}$2 with fundamental
strings of which transverse length is zero in the coincidence
limit of the D2${\bar {\rm D}}$2, i.e., $X^{I}_{1}=X^{I}_{2}$ in
(\ref{dea})--(\ref{ptx}). When the location of the fundamental
strings coincides with the center of global vortex at the origin
$(r=0)$, existence of the fundamental string is marked by
non-vanishing radial component of electric field $F_{0r}=E_r(r)$
on each D2 (or ${\bar {\rm D}}$2) as shown in Fig.~\ref{fig1}. For
simplicity we turn off other components of the field strength,
$F_{0\theta}=0$ and $F_{r\theta}=0$.

Substitution of these into the determinant of (\ref{acg}) gives the action
\begin{equation}\label{acte}
\frac{S}{\int d^{p-2}x_{\perp}}
= -2{\cal T}_{p}\int dtdrd\theta r V(\tau)
\sqrt{X}\, ,
\end{equation}
where, with $g\equiv {\rm det}g_{\mu\nu}$,
\begin{eqnarray}
{\rm det}\left[g_{\mu\nu}+F_{\mu\nu}+(\partial_\mu\bar{T}\partial_\nu
T+\partial_\nu\bar{T}\partial_\mu T)/2\right]&\equiv &g X
\label{detd}\\
&=& -r^2\left[\left(1-E_{r}^{2}+\tau^{'2}\right)
\left(1+\frac{n^2}{r^2}\tau^2\right)\right].
\label{dete}
\end{eqnarray}
From the action (\ref{acte}),
tachyon equation of motion is given by
\begin{equation}\label{ete}
\frac{1}{r}\frac{d}{dr}\left[
\frac{rV}{\sqrt{X}}\left(1+\frac{\tau^2}{r^2}n^2\right)~\tau'\right]
-\frac{V}{\sqrt{X}}(1+\tau^{'2}-E_r^2)\tau\frac{n^2}{r^2}
=\sqrt{X}~\frac{dV}{d\tau} .
\end{equation}

The gauge field strength $F_{0r}=E_r(r)$ with other components vanishing
automatically satisfies
Bianchi identity,
$\partial_{\mu}F_{\nu\rho}+\partial_{\nu}F_{\rho\mu}
+\partial_{\rho}F_{\mu\nu}=0$.
Then equation for the gauge field
leads to constancy of the conjugate momentum $\Pi^{r}$ multiplied by $r$
\begin{eqnarray}\label{gaeq}
(r\Pi^{r})^{'}=0\quad \mbox{for} \quad r\ne 0,
\end{eqnarray}
where the conjugate momenta for the gauge field $A_{r}$ is
\begin{eqnarray}\label{pir}
\Pi^{r}\equiv \frac{1}{\sqrt{-g}}\frac{\delta S}{\delta
(\partial_{t}A_{r})}=\frac{1}{\sqrt{-g}} \frac{\delta S}{\delta
F_{tr}}= \frac{2{\cal T}_{p}V}{\sqrt
X}\left(1+\frac{n^2}{r^2}\tau^2\right) E_r(r).
\end{eqnarray}
Now we obtain an expression for the electric field from the gauge
equation (\ref{gaeq}) with (\ref{pir})
\begin{eqnarray}\label{er}
E_r^2(r)=\frac{
1+\tau'^2}
{1+\left(\frac{2{\cal T}_{p}V}{\Pi^{r}}\right)^{2}
\left(1+\frac{n^2}{r^2}\tau^2\right)} ,
\end{eqnarray}
and thereby the only equation to solve is the tachyon equation
(\ref{ete}). Note that non-vanishing component of conservation of
the energy-momentum tensor, $\nabla_{\mu}T^{\mu \nu}=0$, is only
the radial component,
\begin{eqnarray} \label{ete2}
\frac{d}{dr}\left[\frac{V}{\sqrt{X}}\left(1+\frac{n^2}{r^2}\tau^{2}\right)
\right]
+\frac{V}{r\sqrt{X}}\left(E_r^2
+\frac{n^2}{r^2}\tau^2-\tau^{'2}\right)=0 ,
\end{eqnarray}
which is consistent with the tachyon equation (\ref{ete}).

Let us try to find asymptotic solution. If the tachyon
amplitude approaches infinity with showing power-law behavior
$\tau\sim \tau_\infty r^k$ ($k>0$), $1-E_r^2+\tau^{'2}$ and $X$
would be exponentially suppressed due to the factor $V^2$ ($\sim
4{\rm exp}[-2\tau_\infty r^k/R]$). When $0<k<1$ and $k>1$, one can
easily notice mismatches between power-law behavior in the
left-hand side of (\ref{ete2}) and exponentially-decreasing
behavior in the right-hand side as follows,
\begin{equation}
\left\{
\begin{array}{cc}
k>1: & \frac{(k-1)}{\tau_{\infty} k}r^{-k}\\[2mm]
0<k<1: &\tau_{\infty} k(k-1) r^{k-2}
\end{array}
\right\} \approx 0~~{\rm exponentially} ,
\end{equation}
Neglecting $V^2$, the tachyon equation Eq.~(\ref{ete})
becomes much simpler,
\begin{eqnarray}
\frac{d}{dr}\left[\frac{\tau^{'}rV}{\sqrt{X}} \left(1+\frac{n^2
\tau^2}{r^2}\right)\right] \approx 0 ,
\end{eqnarray}
and the solution of constant $\tau^{'}$ with $\tau\sim \tau_\infty r$
is allowed as a leading behavior. Specifically,
$\tau (r)$ and $E_r^2(r)$ are
\begin{eqnarray} \label{solT}
\tau (r)&\approx& \tau_\infty r + \delta  -
\frac{4{\cal T }_p^2R}{\tau_\infty^2Q_{\rm
F1}^2}(1+\tau_\infty^2)(1+n^2\tau_\infty^2)
r^2e^{-2\frac{\tau_\infty r+\delta}{R}} +{\cal
O}(re^{-2\frac{\tau_\infty r+\delta}{R}}),
\\ \label{solE}
E_r^2(r)&\approx&
(1+\tau_\infty^2)\left\{1-\frac{16{\cal
T}_p^2R}{\tau_\infty Q_{\rm
F1}^2}\left[1+n^2\tau_\infty^2\left(1+\frac{2\delta}{R}\right)
\right]re^{-2\frac{\tau_\infty
r+\delta}{R}} + {\cal O}(e^{-2\frac{\tau_\infty
r+\delta}{R}})\right\} ,
\end{eqnarray}
where the constants, $\tau_\infty$ ($>0$) and $\delta$,
cannot be determined by the information at asymptotic region alone.
Unlike the case of single topological BPS tachyon
kink~\cite{Kim:2003in}, the electric field does not approach
critical value but a different value in order to make the
determinant (\ref{dete}) vanish at infinity.

Near the origin consistent power series expansion leads to increasing
$\tau$ and decreasing $E_{r}^{2}$ as
\begin{eqnarray}
\tau(r)&\approx &\tau_0 r\left[1 + \frac{2{\cal T}_p^{2}}{3Q_{{\rm
F}1}^{2}}(1+\tau_0^2)(n^2-1)r^2
- {\cal O}(r^4)\right] , \label{tr00}\\
E_{r}^{2}(r) &\approx & (1+\tau_0^2)\left[ 1-\frac{4{\cal
T}_p^{2}}{Q_{{\rm F}1}^{2}}(1+\tau_0^2) r^2 +{\cal O}(r^4)\right]
, \label{ee0}
\end{eqnarray}
where $Q_{{\rm F}1}$ is charge of the fundamental string from (\ref{gaeq})
\begin{equation}\label{fcd}
\Pi^{r}\equiv \frac{Q_{{\rm F1}}}{r}.
\end{equation}

\begin{figure}[t]
\centerline{\epsfig{figure=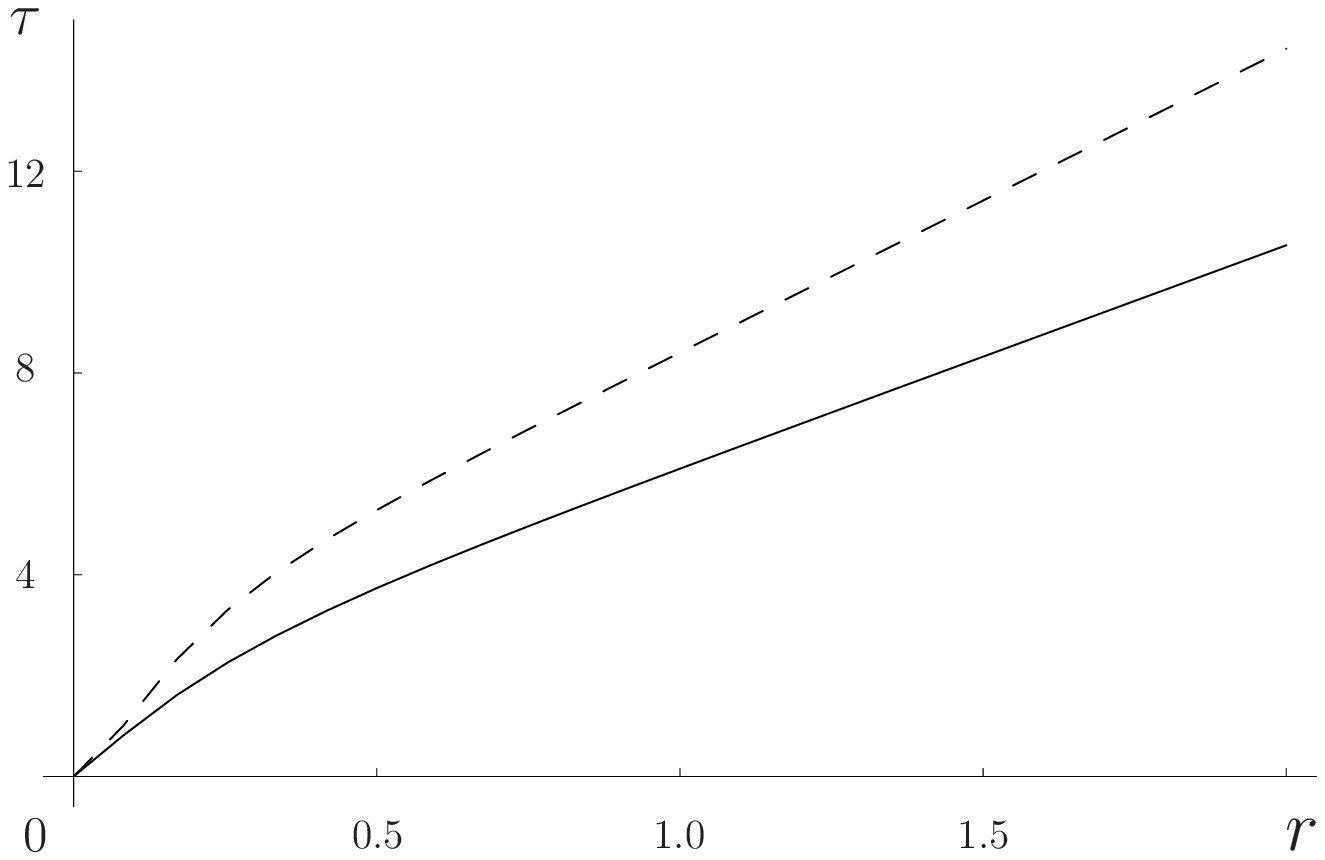,height=6cm}}
\centerline{(a)}
\centerline{\epsfig{figure=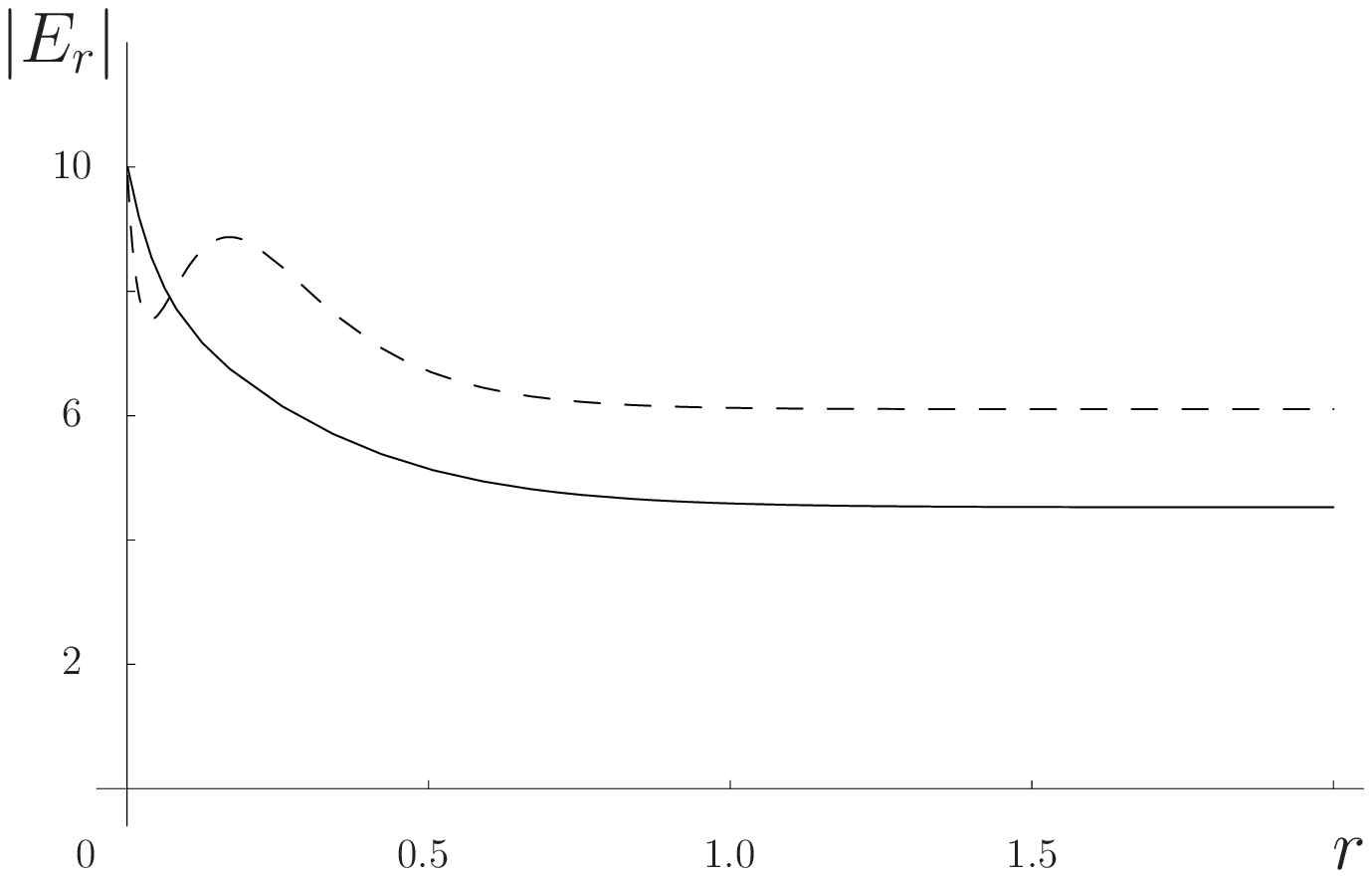,height=6cm}}
\centerline{(b)}
\caption{(a) $\tau(r)$. (b) $|E_{r}(r)|$. $n=1$ case for solid lines
and $n=2$ case for dashed lines. We set $2{\cal T}_{p}=Q_{{\rm F}1}$ for
numerical analysis.}
\label{fig2}
\end{figure}

Numerical solution connecting the behavior near the origin
(\ref{tr00}) and that for large $r$ (\ref{solT}) is shown in
Fig.~\ref{fig2}-(a). Since the coefficient of $r^{3}$-term in
(\ref{tr00}) is positive for $n\ge 2$ but that of $r^{5}$-term in
(\ref{tr00}) is negative for $n=1$, $\tau$ for $n=2$ is convex-up
near the origin but that for $n=1$ is convex-down (See
Fig.~\ref{fig2}-(a)). Though the string charge density $\Pi^{r}$
(\ref{fcd}) has singularity at the origin, the corresponding
electric field $E_{r}$ is regular as shown in (\ref{ee0}) (See
Fig.~\ref{fig2}-(b)). From the graphs Fig.~\ref{fig2}-(a) and (b),
we read the values of $\tau_{\infty}$, $\tau_{\infty}=4.41$ for
$n=1$ and $\tau_{\infty}=6.02$ for $n=2$, which satisfy the
consistency condition $E_{r}^{2}(\infty) \approx
1+\tau^{2}(\infty)$. The leading behavior of tachyon is dominated
by linear term near the origin (\ref{solT}) and at infinity
(\ref{tr00}) irrespective of the vorticity $n$ although values of
the coefficients are different, i.e., $\tau_{0}\ne \tau_{\infty}$
from Fig.~\ref{fig2}-(a). This phenomenon is different from that
of the vortices in local field theory models, and possibly is a
reflection of the string theory: In boundary string field theory,
D0-brane obtained from D2${\bar {\rm D}}$2-system has been
analyzed by employing a linear configuration of which BPS limit is
achived in the limit of infinite slope~\cite{Kraus:2000nj}.

Energy density $-T^{t}_{\;t}$ and radial pressure $T^{r}_{\;r}$ have
the same functional forms with (\ref{fttt})--(\ref{ftrr}) by replacing
$X$ in (\ref{dete}), but angular pressure is different;
\begin{eqnarray}\label{teth}
T^{\theta}_{\; \theta}=-\frac{2{\cal T}_{p}V}{\sqrt{X}}
\left(1+\tau'^{2}-E_{r}^{2}\right).
\end{eqnarray}
$\theta$-component of the U(1) current is
\begin{eqnarray}\label{fjh}
j^{\theta}=\frac{2{\cal T}_{p}V}{\sqrt{X}}(1+\tau'^{2}-E_{r}^{2})
\frac{n}{r^{2}}\tau^{2} ,
\end{eqnarray}
and it satisfies conservation law $\nabla_{\mu}j^{\mu}=0$.

Near the origin we read behavior of densities, (\ref{fttt})--(\ref{ftrr}) and
(\ref{teth})--(\ref{fjh}),
by substituting (\ref{tr00})--(\ref{ee0})
\begin{eqnarray}
T^{t}_{\; t}&\approx & -\sqrt{1+\tau_{0}^{2}} \,\frac{|Q_{{\rm
F1}}|}{r}\left\{1+\frac{2{\cal T }_p^2}{Q_{\rm
F1}^2}~[1+\tau_0^2(2n^2-1)] r^2 +{\cal O}(r^4)\right\},
\label{et0}\\
T^{r}_{\; r} &\approx & -\frac{1}{\sqrt{1+\tau_{0}^{2}} }
\,\frac{|Q_{{\rm F1}}|}{r}\left\{1 + \frac{2{\cal T}_p^2}{Q_{\rm
F1}^2}(1+\tau_0^2)~ r^2+{\cal O}(r^4)\right\},
\label{er0}\\
T^{\theta}_{\; \theta}&\approx & -4{\cal
T}_{p}^2\sqrt{1+\tau_{0}^{2}}\, \frac{r}{|Q_{{\rm F1}}|}+{\cal O}(r^3),\\
j^{\theta}&\approx &  4{\cal
T}_{p}^2\tau_0^2\sqrt{1+\tau_{0}^{2}}\, \frac{r}{|Q_{{\rm
F1}}|}+{\cal O}(r^3).
\end{eqnarray}
The energy density (\ref{et0}) and radial pressure (\ref{er0}) are
singular at the origin.
Here the source of the singularity
is identified with the singular fundamental string charge (\ref{fcd})
at the origin so that contribution from the vortex configuration
represented by the tachyon amplitude is regular.
The singular behavior in this DBI type theory becomes milder
(${\cal O}(1/r)$) in comparison with that in Maxwell theory in a
plane ($E_{r}^{2}\sim 1/r^{2}$) so that self-energy of the
fundamental string does not
involve UV divergence.

From (\ref{solT}) and (\ref{solE}),
the asymptotic forms of energy density and radial pressure
show long-range term for sufficiently large $r$, but the angular
pressure and the U(1) current do not as expected;
\begin{eqnarray}
T^{t}_{\; t}&\approx & -\sqrt{1+\tau_{\infty}^{2}}
\,~\frac{|Q_{{\rm F1}}|}{r}\left[1+\frac{16{\cal
T }_p^2}{Q_{\rm
F1}^2}(1+n^2\tau_\infty^2)r^2e^{-2\frac{\tau_\infty
r+\delta}{R}}+{\cal O }(re^{-2\frac{\tau_\infty
r+\delta}{R}})\right],
\label{eti}\\
T^{r}_{\; r} &\approx & -\frac{1}{\sqrt{1+\tau_{\infty}^{2}} }
\,~\frac{|Q_{{\rm F1}}|}{r}\left\{1+\frac{8{\cal T
}_p^2R}{\tau_\infty Q_{\rm
F1}^2}\left[1+n^2\tau_\infty^2\left(1+\frac{2\delta}{R}\right)\right]
re^{-2\frac{\tau_\infty
r+\delta}{R}}\right. \nonumber\\
&&\hspace{37mm}\left. +{\cal O }(e^{-2\frac{\tau_\infty
r+\delta}{R}})\right\},
\label{eri}\\
T^{\theta}_{\; \theta}&\approx & -16{\cal
T}_{p}^2\frac{\sqrt{1+\tau_{\infty}^{2}}}{|Q_{\rm
F1}|}re^{-2\frac{\tau_\infty r+\delta}{R}}\left[1+{\cal
O}(1/r)\right],
\label{thhi}\\
j^{\theta}&\approx &  16{\cal
T}_{p}^2n\frac{\tau_{\infty}^2\sqrt{1+\tau_{\infty}^{2}}}{|Q_{\rm
F1 }|}re^{-2\frac{\tau_\infty r+\delta}{R}}\left[1+{\cal
O}(1/r)\right].
\label{jti}
\end{eqnarray}
Therefore, energy of the obtained vortex configuration is linearly divergent
\begin{equation}
E=2\pi\int_0^{R_{{\rm IR}}}dr r T_{tt}\propto Q_{{\rm F1}}R_{{\rm
IR}} \stackrel{R_{{\rm IR}}\rightarrow \infty}{\longrightarrow }\infty
.
\end{equation}
This linear divergence due to the fundamental string charge at the origin
might be different from the familiar nature of
logarithmically divergent energy of the global vortex.

\begin{figure}[t]
\centerline{\epsfig{figure=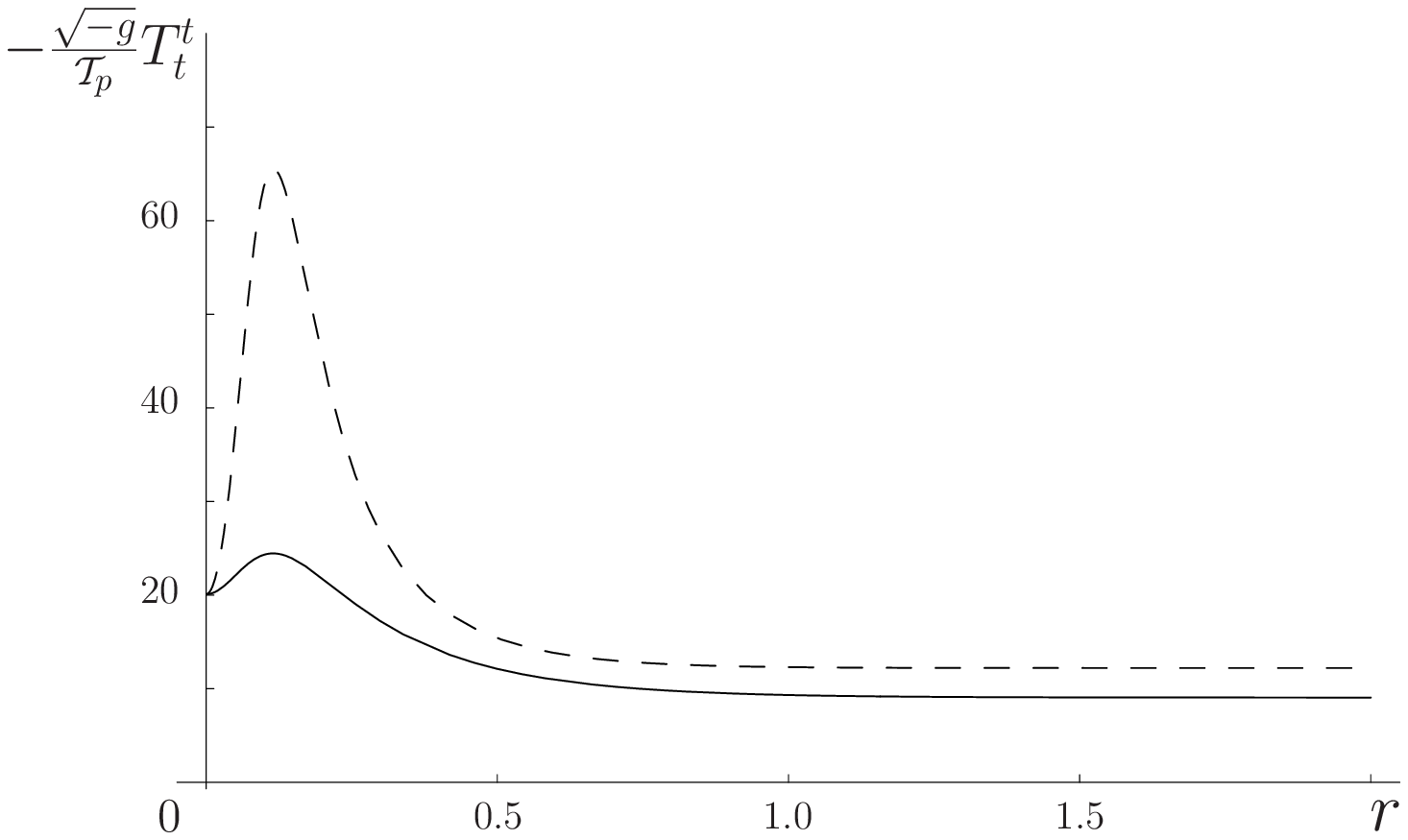,height=6.2cm}}
\centerline{(a)}
\centerline{\epsfig{figure=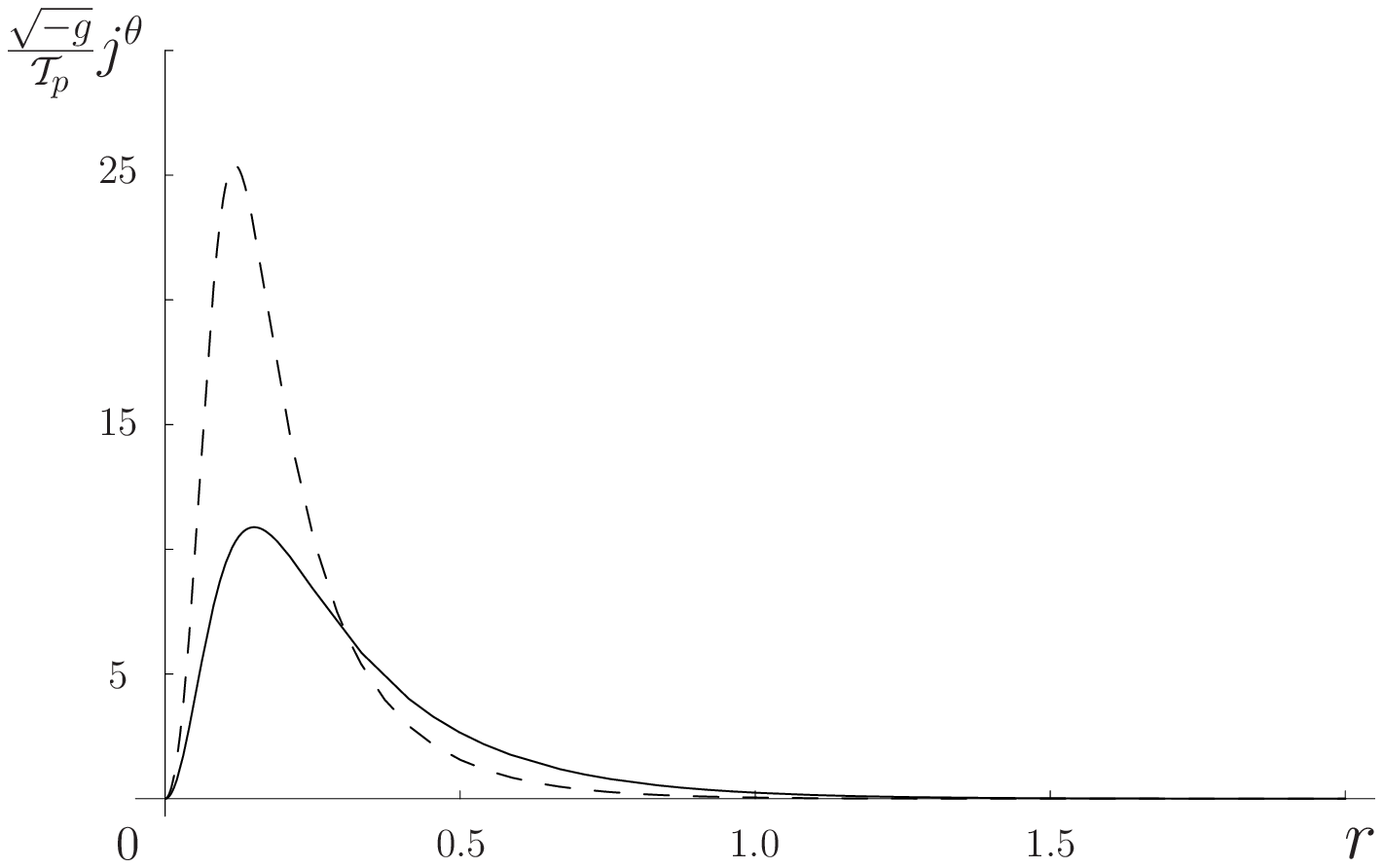,height=6.2cm}}
\centerline{(b)}
\caption{(a) $-\sqrt{-g}\, T^{t}_{\;t}/{\cal T}_{p}$.
(b) $\sqrt{-g}\, j^{\theta}/{\cal T}_{p}$.
$n=1$ case for solid lines and $n=2$ case for dashed lines.}
\label{fig3}
\end{figure}

Radial distribution of $-\sqrt{-g}\, T^{t}_{\;t}$ and $-\sqrt{-g}\, j^{\theta}$
are drawn in Fig.~\ref{fig3}. From Fig.~\ref{fig3}-(a) we can easily read
the numerical values of $\tau_{0}$ under $2{\cal T}_{p}=Q_{{\rm F}1}$
and $\tau_{\infty}$, and the U(1) current
is localized near the origin as shown in Fig.~\ref{fig3}-(b).
Thin limit is achieved by taking the limit of infinite $\tau_\infty$.
Then, from (\ref{solE}) and (\ref{eti})--(\ref{jti}), $E_{r}$ and
$-\sqrt{-g}\, T^{t}_{\;t}$ go to infinity but all the pressure components
and the U(1) current vanish, which lets the limiting process of
singular zero thickness
with keeping constant $Q_{{\rm F}1}$ smooth.

\setcounter{equation}{0}
\section{Local D-Vortices}
When the U(1)$\times$U(1) gauge fields in (\ref{dea}) are
rewritten as $A_{\mu}=(A^{1}_{\mu}+A^{2}_{\mu})/2$ and $C_{\mu}=(A^{1}_{\mu}
-A^{2}_{\mu})/2$, the former $A_{\mu}$ remains to be massless in
symmetric phase and the latter $C_{\mu}$ becomes massive in broken phase.
It is easily read by the form of the action
\begin{eqnarray}\label{acg1}
S=-{\cal T}_{p}\int d^{p+1}x\, V(T)\left[\,\sqrt{-\det(X^{+}_{\mu\nu})}
+\sqrt{-\det(X^{-}_{\mu\nu})}\,\,\right],
\end{eqnarray}
where $C_{\mu\nu}=\partial_{\mu}C_{\nu}-\partial_{\nu}C_{\mu}$,
$D_{\mu}T=(\partial_{\mu} -2iC_{\mu})T$, and
\begin{equation}\label{Xpm}
X^{\pm}_{\mu\nu}=g_{\mu\nu}+F_{\mu\nu}\pm C_{\mu\nu}
+({\overline {D_{\mu}T}}D_{\nu}T +{\overline {D_{\nu}T}}D_{\mu}T)/2 .
\end{equation}
Note that the charge of $T$ is 2 in the unit system of
consideration. In this coincidence limit, the D-brane is distinguished
from the $\bar{{\rm D}}$-brane by coupling to the gauge field $C_{\mu}$.

For $F_{\mu\nu}$, we employ the same
configuration, i.e., all the components vanish for singular local D-vortex and
$F_{0r}$ is turned on for regular local D-vortex as has been done for global
D-vortices.

\subsection{Singular local D-vortex}

In the subsection 2.1, we dealt with global D-vortices with vanishing
$F_{\mu\nu}$. In this subsection we consider local D-vortices with
vanishing $F_{\mu\nu}$ but with nonvanishing gauge field $C_{\mu}$.
A minimal gauge field configuration for the local D-vortices
superimposed at the origin is
to turn only on the angular component
$C_{\mu}=\delta_{\mu}^{\;\theta}C_{\theta}$ under the Weyl gauge $C_{0}=0$.

Substituting the ansatz (\ref{ans}) and the gauge field $C_{\theta}$
into the equations of motion, we obtain tachyon equation
\begin{eqnarray}\label{setel}
\frac{1}{r}\frac{d}{dr}\left\{
\frac{rV}{\sqrt{X}}\left[1+\frac{\tau^2}{r^2}
(n-2C_\theta)^2\right]~\tau'\right\}
-\frac{V}{\sqrt{X}}(1+\tau^{'2})\tau\frac{(n-2C_\theta)^2}{r^{2}}
=\sqrt{X}~\frac{dV}{d\tau} ,
\end{eqnarray}
and equation of the gauge field
\begin{eqnarray}\label{sceq}
\frac{d}{dr}\left(\frac{rV}{\sqrt{X}}\frac{C_{\theta}^{'}}{r^{2}}\right)
+2\frac{rV}{\sqrt{X}}\left(1+\tau'^2\right)
\frac{\tau^2}{r^2}(n-2C_{\theta})=0,
\end{eqnarray}
where $C_{r\theta}=C_{\theta}^{'}=dC_{\theta}/dr$ and
\begin{equation}\label{siX}
X\equiv \left(1+\tau^{'2}\right)\left[
1+(n-2C_{\theta})^{2}\frac{\tau^{2}}{r^{2}}\right]+
\frac{C_{\theta}^{'2}}{r^{2}}.
\end{equation}

Since we look for nonsingular solutions of the equations of
motion, (\ref{setel}) and (\ref{sceq}), the ansatz
(\ref{ans}) is used for the tachyon field and the form of gauge field,
$C_{i}=-\epsilon_{ij}x^{j} C_{\theta}(r)/r^{2}$
($\epsilon_{12}=1$), forces the following boundary condition at the origin as
\begin{equation}\label{bd0l}
C_{\theta}(r=0)=0.
\end{equation}
In addition to the boundary condition of the tachyon field at infinity
(\ref{bdin}), value of the gauge field $C_{\theta}$ at infinity should be
\begin{equation}\label{cdi}
C_{\theta}(r=\infty)=n/2
\end{equation}
to make the $(n-2C_{\theta})$-term in the energy density,
radial pressure, and $\theta$-component of U(1) current vanish at infinity;
\begin{eqnarray}
T^{t}_{\; t}&=&-\frac{2{\cal T}_{p}V}{\sqrt{X}}
\left\{\left(1+\tau'^{2}\right)\left[1+(n-2C_{\theta})^{2}
\frac{\tau^2}{r^2}\right] +\frac{C_{\theta}^{'2}}{r^{2}}\right\},
\label{ftttl}\\
T^{r}_{\; r}&=&-\frac{2{\cal T}_{p}V}{\sqrt{X}}
\left[1+(n-2C_{\theta})^{2}\frac{\tau^{2}}{r^{2}}\right],
\label{ftrrl}\\
j^{\theta}&=&\frac{2{\cal T}_{p}V}{\sqrt{X}}(1+\tau'^{2})
\frac{\tau^{2}}{r^{2}}(n-2C_{\theta}).
\label{fjhl}
\end{eqnarray}
Another nonvanishing component of energy-momentum tensor is
\begin{eqnarray}\label{ftth}
T^{\theta}_{\; \theta}&=&-\frac{2{\cal T}_{p}V}{\sqrt{X}}
\left(1+\tau'^{2}\right).
\end{eqnarray}
If we try $C_\theta(r)\approx C_\infty r^q$ to the equation of gauge field,
its left-hand side of (\ref{sceq}) vanishes only when (\ref{cdi}) is satisfied.

Substituting the boundary conditions at infinity, (\ref{bdin}) and (\ref{cdi}),
into the tachyon equation, (\ref{setel}) reduces approximately to
the tachyon equation with zero vorticity $(n=0)$ and gauge field
$(C_\theta =0)$:
\begin{eqnarray}
\tau''+\left(\frac{1}{r}-\frac{\tau'}{R}-\frac{1}{2}\frac{X'}{X}
\right)\tau' \approx -\frac{1+\tau^{'2}}{R}.
\end{eqnarray}
For the tachyon $\tau(r)\approx \tau_\infty r^k,~(k\geq 0)$ at
asymptotic region, the left-hand side cannot be equated with the
right-hand side. Therefore, nonexistence of static regular local
D-vortex solutions satisfying (\ref{bdin}) and (\ref{cdi}) is
obvious. It means that the static singular local D-vortex
configuration in Ref.~\cite{Sen:2003tm} cannot be understood as a
singular limit of static regular local D-vortex solution.

We showed nonexistence of the regular static local vortices satisfying the
boundary conditions, (\ref{bd0})--(\ref{bdin}) and
(\ref{bd0l})--(\ref{cdi}), however it has been known to exist the singular
local vortices satisfying the same boundary conditions in
Ref.~\cite{Sen:2003tm}. Let us reconfirm the existence of
such solution relying on
the information of boundary string field theory in what
follows~\cite{Kraus:2000nj}.

Suppose that the vortex configuration of vorticity $n$ is
given by $\tau\sim r/\epsilon$ as has been studied
in boundary string field theory~\cite{Kraus:2000nj} and in the
subsection 2.1. For the gauge field we set $C_\theta = (n/2)[1-f(r)]$,
where $f(r)$ should satisfy $f(0)=1$ and $f(\infty)=0$ according to
the boundary conditions (\ref{bd0l}) and (\ref{cdi}).
Substituting these into (\ref{siX}), we have
\begin{eqnarray} \label{x}
X=\left(1+\frac{1}{\epsilon^2}\right)
\left(1+\frac{n^2}{\epsilon^2}f^2\right)
+\frac{n^2}{4r^2}f'^{2} .
\end{eqnarray}
Now let us assume that $\epsilon$ is sufficiently small. Since $f(r)$
is a monotonic decreasing function, $f$ has almost unity for
$r<\epsilon R$ and zero for $r>\epsilon R$. In the expression of $X$ (\ref{x}),
possible candidates of the leading term would be $1/\epsilon^{2}$,
$n^{2}f^{2}/\epsilon^{4}$, and
$n^{2}f'^{2}/4r^{2}$. Near $r\approx 0$, $n^{2}f^{2}/\epsilon^{4}$ or
$n^{2}f'^{2}/4r^{2}$ is dominant so that (\ref{sceq}) supports a solution
$f\sim e^{-\frac{4R}{3\epsilon^3}r^3}$. For large $r$, $1/\epsilon^{2}$-term
is dominant so that the approximated equation of (\ref{sceq}) leads to
$f(r)\approx ce^{-r^2/\epsilon^2}$. When $\epsilon\rightarrow 0$,
the energy density has $\delta$-function like singularity at the origin,
$-T^{t}_{\;t}(0)\approx 2{\cal T}_{p}n/\epsilon^{2}\stackrel{\epsilon
\rightarrow 0}{\longrightarrow \infty}$, but the pressures remain constant,
$T^{r}_{\;r}(0)\approx -2{\cal T}_{p}n$ and
$T^{\theta}_{\;\theta}(0)\approx -2{\cal T}_{p}/n$. At $r\ne 0$,
$-T^{t}_{\;t}=T^{r}_{\;r}=T^{\theta}_{\;\theta}=0$ for $n=1$.
Therefore, the conservation law equivalent to the tachyon equation is
satisfied in the $\epsilon\rightarrow 0$ limit, i.e., $(T^{r}_{\; r})'
=-(T^{r}_{\; r} -T^{\theta}_{\; \theta})/r =0$.

Tension of the singular D$0$-brane with vanishing thickness is
\begin{eqnarray}
2{\cal T}_0^{\ell}&=&\int_0^\infty d\theta\int_0^\infty dr r(-T^t_t)\nonumber\\
&\approx& 4\pi R^{2}{\cal T}_2 \int_0^\infty dy
\frac{\sqrt{1+16y^{2}} e^{-16y^{3}/3}}{\cosh y},\\
&=& 2(\sqrt{2\times 0.26254\times \pi} R)^{2}{\cal T}_{2},
\label{tel}
\end{eqnarray}
where $y=r/\epsilon R$ and $R=\sqrt{2}$.
If we compare the obtained tension of local singular D-vortex (\ref{tel}) with
that of global singular D-vortex (\ref{cata}), the energy of singular
D-vortex is significantly lowered by the gauge field $C_{\theta}$,
i.e., ${\cal T}_{0}^{\ell}/{\cal T}_{0}^{g}$ is about 0.14.
Due to the gauge field, the D$0$-brane also carries the quantized
magnetic flux
\begin{eqnarray}
\Phi&=&\int_{0}^{\infty} dr\int_{0}^{2\pi}d\theta \,
C_{r\theta}\nonumber\\
&=&2\pi \left[C_{\theta}(r=\infty)-C_{\theta}(r=0)\right]\\
&=&\pi n.
\label{flu}
\end{eqnarray}

\subsection{Regular local D-vortex with electric flux}

Let us turn on $F_{0r}=E_{r}$ again.
Then the effective action (\ref{acg1}) becomes
\begin{equation}\label{actr1}
\frac{S}{\int d^{p-2}x_{\perp}}
= -2{\cal T}_{p}\int dtdrd\theta r V(\tau)\sqrt{X}\, ,
\end{equation}
where
\begin{equation}\label{Xl}
X\equiv \left(1+\tau^{'2}-E_{r}^{2}\right)\left[
1+(n-2C_{\theta})^{2}\frac{\tau^{2}}{r^{2}}\right]+
\frac{C_{\theta}^{'2}}{r^{2}}.
\end{equation}

We read tachyon equation
\begin{eqnarray}\label{etel}
\frac{1}{r}\frac{d}{dr}\left\{
\frac{rV}{\sqrt{X}}\left[1+
\frac{\tau^2}{r^2}(n-2C_\theta)^2\right]~\tau'\right\}
-\frac{V}{\sqrt{X}}(1+\tau^{'2}-E_r^2)\tau\frac{(n-2C_\theta)^2}{r^{2}}
=\sqrt{X}~\frac{dV}{d\tau} .
\end{eqnarray}
Equation for the gauge field $A_{\mu}$ is obtained from
constancy of the conjugate momentum $\Pi^{r}$ multiplied by $r$ (\ref{gaeq}).
Specific form of the conjugate momentum $\Pi^{r}$ is
\begin{eqnarray}\label{pirl}
\Pi^{r}\equiv \frac{1}{\sqrt{-g}}\frac{\delta S}{\delta
(\partial_{t}A_{r})}=\frac{1}{\sqrt{-g}} \frac{\delta S}{\delta
F_{tr}}= \frac{2{\cal T}_{p}V}{\sqrt
X}\left[1+(n-2C_{\theta})^{2}\frac{\tau^2}{r^2}\right] E_r(r),
\end{eqnarray}
and thereby the electric field is
\begin{eqnarray}\label{erl}
E_r^2(r)=\frac{
(1+\tau'^2)\left[1+(n-2C_{\theta})^{2}\frac{\tau^2}{r^2}\right]
+\frac{C_{\theta}^{'2}}{r^{2}}}
{\left[1+(n-2C_{\theta})^{2}\frac{\tau^2}{r^2}\right]
\left\{1+\left(\frac{2{\cal T}_{p}V}{\Pi^{r}}\right)^{2}
\left[1+(n-2C_{\theta})^{2}\frac{\tau^2}{r^2}\right]\right\}} .
\end{eqnarray}
Nonvanishing field strength of the gauge field $C_{\mu}$ is
$C_{r\theta}=C_{\theta}^{'}(r)$
which automatically satisfies Bianchi identity,
$\partial_{\mu}C_{\nu\rho}+\partial_{\nu}C_{\rho\mu}
+\partial_{\rho}C_{\mu\nu}=0$.
Equation of the gauge field $C_{\mu}$ is
\begin{eqnarray}\label{ceq}
\frac{1}{r}\frac{d}{dr}\left(\frac{rV}{\sqrt{X}}
\frac{C_{\theta}^{'}}{r^{2}}\right)
+2\frac{V}{\sqrt{X}}\left(1+\tau'^2 -E_{r}^{2}\right)\tau^2
\frac{n-2C_{\theta}}{r^{2}}=0.
\end{eqnarray}

Regarding the boundary condition of radial component of electric field
$E_{r}$ at infinity, vanishing $\theta$-component of pressure
\begin{eqnarray}\label{ftthe}
T^{\theta}_{\; \theta}&=&-\frac{2{\cal T}_{p}V}{\sqrt{X}}
\left(1+\tau'^{2}-E_{r}^{2}\right)
\end{eqnarray}
requires
\begin{equation}\label{edi}
E_{r}^{2}(\infty)\rightarrow 1+\tau^{'2}(\infty).
\end{equation}
Then the $\theta$-component of U(1) current
\begin{equation}\label{fjhle}
j^{\theta}=\frac{2{\cal T}_{p}V}{\sqrt{X}}\left(1+\tau'^{2}-E_{r}^{2}\right)
\frac{\tau^{2}}{r^{2}}(n-2C_{\theta})
\end{equation}
also decays to zero rapidly as $r$ increases.
Note that the energy density $-T^{t}_{\;t}$
and the radial pressure $T^{r}_{\;r}$
with nonvanishing $E_{r}$ share the same functional forms with
$T^{t}_{\;t}$ (\ref{ftttl}) and $T^{r}_{\;r}$ (\ref{ftrrl})
without $E_{r}$ except for
the difference in the expression of determinant $X$ (\ref{Xl}).

The D-vortices of interest are characterized by the fundamental string
charge $Q_{{\rm F1}}$ with density $\Pi^{r}$ (\ref{pirl})
and the topological charge represented by the magnetic flux
(\ref{flu}).
In the last line (\ref{flu}), we used the boundary
condition (\ref{cdi}), and the result (\ref{flu}) means that unit of
the magnetic flux is $\pi$ where the denominator 2 in $2\pi n/2$ comes from
the charge of complex tachyon field.

Power series expansion near the origin shows that the tachyon $\tau$
and the gauge field $C_{\theta}$ are increasing but
radial electric field $E_{r}$ decreases due to negative $\gamma$;
\begin{eqnarray}
\tau(r)&\approx&\tau_{0}r[1+\alpha r^2+{\cal O}(r^4)],
\label{tao}\\
C_{\theta}(r)&\approx&C_{0}r^{3}[1+\beta r^2+{\cal O}(r^3)],
\label{cto}\\
E_{r}^{2}(r)&\approx& (1+\tau_{0}^2)[1+\gamma r^2+{\cal O}(r^4)],
\label{ero}
\end{eqnarray}
where
\begin{eqnarray}
&&\alpha=-\frac{n^2-1}{6(1+\tau_0^2n^2)}\left[9C_0^2-\frac{4{\cal
T }_p^2}{Q_{\rm F1}^2}(1+\tau_0^2)(1+\tau_0^2n^2)^2\right] ,
\\
&&\beta =\frac{3+\tau_0^2
n^2(5-2n^2)}{10(1+\tau_0^2n^2)^3}\left[9C_0^2-\frac{4{\cal T
}_p^2}{Q_{\rm F1}^2}(1+\tau_0^2)(1+\tau_0^2n^2)^2\right] ,
\\
&&\gamma=\frac{1}{(1+\tau_0^2n^2)^2}\left[9C_0^2-\frac{4{\cal T
}_p^2}{Q_{\rm F1}^2}(1+\tau_0^2)(1+\tau_0^2n^2)^2\right].
\end{eqnarray}
For sufficiently large $r$, functional forms of the tachyon amplitude
and electric field rapidly approach asymptotic configurations
\begin{eqnarray}
\tau(r)&\sim& \tau_{\infty}r +\delta -4\frac{{\cal
T}_p^2R(1+\tau_\infty^2)}{\tau_\infty^2Q_{\rm
F1}^2}~r^2e^{-2\frac{\tau_\infty r+\delta }{R}}+{\cal O}( r
e^{-2\frac{\tau_\infty r+\delta }{R}}),
\label{tan}\\
E_{r}^{2}(r)&\sim& (1+\tau_{\infty}^2) \left[1-16\frac{{\cal
T}_p^2R}{\tau_\infty Q_{\rm F1}^2}re^{-2\frac{\tau_\infty r+\delta
}{R}}+{\cal O}(r^4e^{-4\frac{\tau_\infty r+\delta }{R}})\right],
\label{ern}
\end{eqnarray}
where $\tau_{\infty}$ and $\delta$ are constants which cannot be
determined only by
the boundary conditions at infinity.
For the gauge field $C_{\theta}$, let us try
\begin{eqnarray}
C_{\theta}(r)\sim \frac{n}{2} + \delta C_\theta ,
\label{ctn}
\end{eqnarray}
then the linearized equation of $\delta C_\theta$ from (\ref{ceq}) reduces to
\begin{eqnarray}
M(t)\frac{d^{2}\delta C_\theta}{dt^{2}}
=-\frac{d}{d\delta C_\theta}U(\delta C_\theta),
\qquad t=\kappa r^{3},
~\left(\kappa=\frac{4{\cal T}_p\tau_\infty}{3|Q_{\rm
F1}|}\sqrt{1+\tau_\infty^2}e^{-\frac{\delta}{R}}\right)
\end{eqnarray}
which describes the one-dimensional motion of a hypothetical point particle
with exponentially-increasing ``time-dependent
mass'' $M(t)=e^{2\tau_\infty t^{1/3}/(R\kappa^{1/3})}$.
Since the force is given by a conservative
potential $U=-(\delta C_\theta)^{2}/2$, possible
boundary values of $\delta C_\theta$ at infinity are read as
$\delta C_\theta(\infty)=\pm\infty~(\mbox{minima})$
or $\delta C_\theta(\infty)=0~(\mbox{maximum})$.
Though we do not obtain the solution satisfying $\delta C_\theta(\infty)=0$
$(C_\theta(\infty)=n/2)$ analytically, existence of such solution is obvious.
The boundary condition at the origin (\ref{bd0l}) and the shape of
$U(\delta C_\theta)$ require that $\delta C_\theta$ is negative and
monotonic increasing to zero.

\begin{figure}
\centerline{\epsfig{figure=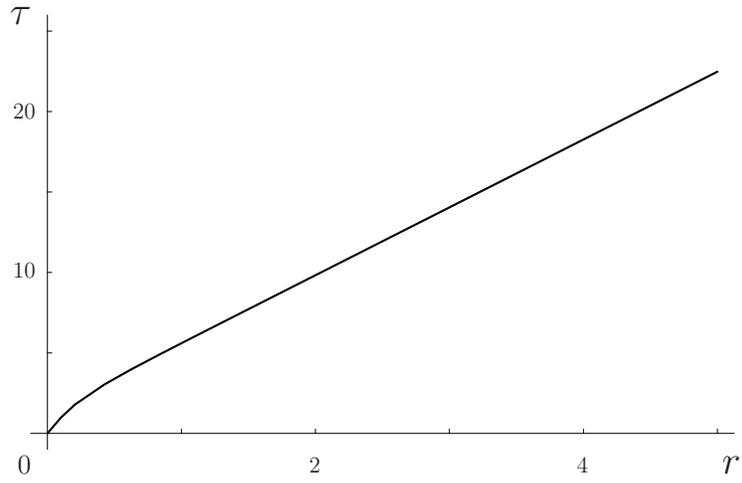,height=6.2cm}}
\centerline{(a)}
\centerline{\epsfig{figure=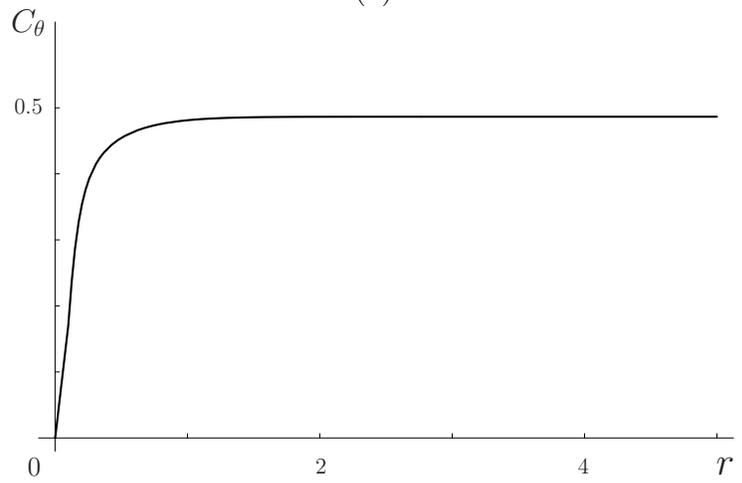,height=6.2cm}}
\centerline{(b)}
\centerline{\epsfig{figure=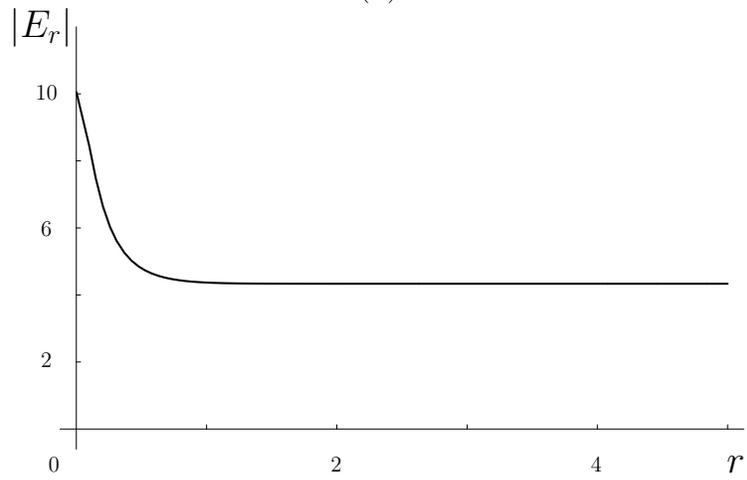,height=6.2cm}}
\centerline{(c)}
\caption{(a) $\tau(r)$. (b) $C_{\theta}(r)$. (c) $|E_{r}(r)|$.
$n=1$ and $2{\cal T}_{p}/Q_{{\rm F}1}=1.2$.}
\label{fig4}
\end{figure}

The obtained numerical results in Fig.~\ref{fig4}
confirm the approximations at both the origin and infinity discussed
previously.
From Fig.~\ref{fig4}-(a) and (c), the numerical value of $\tau_{\infty}$ is
read to be 4.217 under $2{\cal T}_{p}/Q_{{\rm F}1}=1.2$.

\begin{figure}[ht]
\centerline{\epsfig{figure=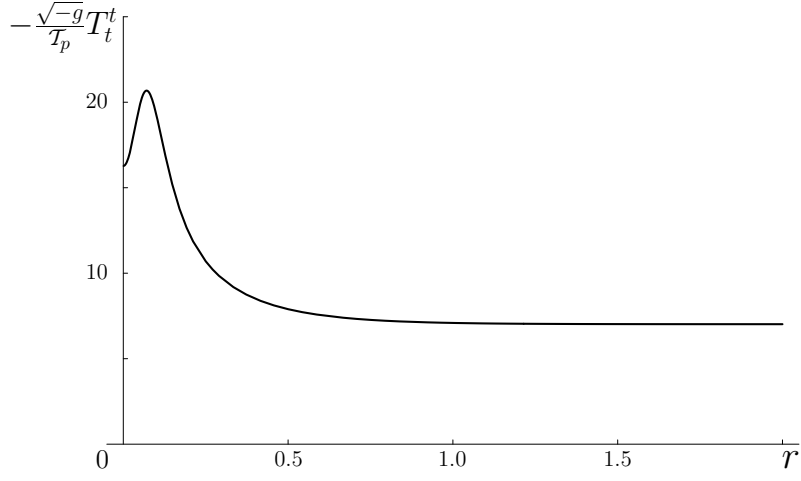,height=6.2cm}}
\centerline{(a)}
\centerline{\epsfig{figure=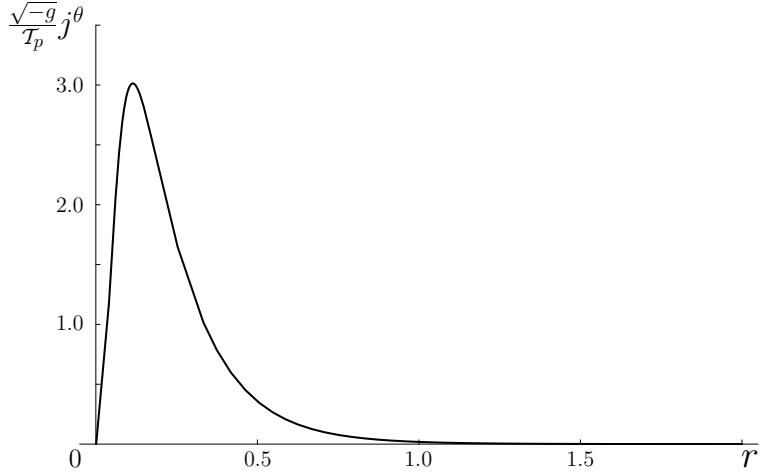,height=6.2cm}}
\centerline{(b)}
\caption{(a) $-\sqrt{-g}\, T^{t}_{\;t}/{\cal T}_{p}$.
(b) $\sqrt{-g}\, j^{\theta}/{\cal T}_{p}$.}
\label{fig5}
\end{figure}

If (\ref{tao})--(\ref{cto}) are inserted in the energy-momentum
tensor, (\ref{ftttl})--(\ref{ftrrl}), (\ref{ftthe}), and the current density
(\ref{fjhle}), then profiles near the origin are
\begin{eqnarray}
T^{t}_{\;t}&\approx & -\sqrt{1+\tau_{0}^2}~\frac{|Q_{\rm
F1}|}{r}\left\{1+\frac{r^2}{2(1+\tau_0^2n^2)^2}\left[9C_0^2+\frac{4{\cal
T }_p^2}{Q_{\rm F1}^2}(1+\tau_0^2(2n^2-1))(1+\tau_0^2n^2
)^2\right]\right. \nonumber\\
&&\hspace{35mm}\left. +{\cal O }(r^4)\right\},
\\
T^{r}_{\;r}&\approx & -\frac{1}{\sqrt{1+\tau_{0}^2}}~\frac{|Q_{\rm
F1}|}{r} \left\{1
-\frac{r^2}{2(1+\tau_0^2n^2)^2}\left[9C_0^2-\frac{4{\cal T
}_p^2}{Q_{\rm F1}^2}(1+\tau_0^2)(1+\tau_0^2n^2)^2\right]
\right. \nonumber\\
&&\hspace{35mm}\left. +{\cal O
}(r^4)\right\},
\\
T^{\theta}_{\;\theta}&\approx & \frac{|Q_{\rm
F1}|r}{(1+\tau_0^2n^2)^2\sqrt{1+\tau_0^2}}\left[9C_0^2-\frac{4{\cal
T }_p^2}{Q_{\rm F1}^2}(1+\tau_0^2)(1+\tau_0^2n^2)^2\right]+{\cal O
}(r^3),
\\
j^{\theta}&\approx & -\frac{\tau_0^2n|Q_{\rm
F1}|r}{(1+\tau_0^2n^2)^2\sqrt{1+\tau_0^2}}\left[9C_0^2-\frac{4{\cal
T }_p^2}{Q_{\rm F1}^2}(1+\tau_0^2)(1+\tau_0^2n^2)^2\right]+{\cal O
}(r^3).
\end{eqnarray}
Substituting (\ref{tan})--(\ref{ctn}) into (\ref{ftttl}) (\ref{ftrrl})
(\ref{ftthe}) (\ref{fjhle}),
we notice that, as in the case of global DF-vortices, the energy
density and the radial component of pressure have a long range term
but the $\theta$-components of pressure and U(1) current do not;
\begin{eqnarray}
T^{t}_{\;t}&\sim & -\sqrt{1+\tau_{\infty}^2}~\frac{|Q_{\rm
F1}|}{r} \left[1+\frac{16{\cal T}_p^2}{Q_{\rm
F1}^2}~r^2e^{-2\frac{\tau_\infty r+\delta }{R}}+{\cal
O}(re^{-2\frac{\tau_\infty r+\delta }{R}})\right],
\label{etil}\\
T^{r}_{\;r}&\sim & -\frac{1}{\sqrt{1+\tau_{\infty}^2}}~\frac{|Q_{\rm F1}|}{r}
\left[1+\frac{8{\cal T}_p^2R}{\tau_\infty Q_{\rm F1}^2}
re^{-2\frac{\tau_\infty r+\delta }{R}}
+{\cal O}(r^4e^{-4\frac{\tau_\infty r+\delta }{R}})\right],
\label{eril}\\
T^{\theta}_{\;\theta}&\sim &  -16\frac{{\cal
T}_p^2\sqrt{1+\tau_\infty^2}}{|Q_{\rm
F1}|}~re^{-2\frac{\tau_\infty r+\delta }{R}}\left[1
+{\cal O}(re^{-2\frac{\tau_\infty r+\delta }{R}})\right],\label{thhil}\\
j^{\theta}&\sim & -32{\cal
T}_p^2\frac{\tau_\infty^2\sqrt{1+\tau_\infty^2}}{|Q_{\rm
F1}|}re^{-2\frac{\tau_\infty r +\delta}{R}}\delta C_{\theta}.
\label{jtil}
\end{eqnarray}
Note that the coefficients of the leading terms of $T^{t}_{\;t}$ (\ref{etil})
and $T^{r}_{\;r}$ (\ref{eril}) coincide exactly with those in
(\ref{eti}) and (\ref{eri}) except for the value of $\tau_{\infty}$,
which means that the leading behavior
at asymptotic region is governed by the fundamental strings.
The effect of the gauge field $C_{\theta}$ appears in the subleading term,
which makes the fields approach their boundary behaviors at
infinity more rapidly.
Fig.~\ref{fig5}-(a),(b) show that the localized parts near the origin have
the ring shape. Since the leading term of the energy density
decreases with $1/r$, we can read the value of $\tau_{\infty}=
4.217$ again from the Fig.~\ref{fig5}-(a), which is exactly the same as that
from the figure of $\tau(r)$ in Fig.~\ref{fig2}-(a).

Though we obtained the regular solutions in subsections 2.2 and 3.2,
the stability of D-vortices and D-strings with electric flux should be
studied including production of closed string degrees of
freedom~\cite{Lambert:2003zr}
as has been discussed in Ref.~\cite{Leblond:2004uc}.

\setcounter{equation}{0}
\section{From D-vortices to D-strings}

For unstable D-branes, the coupling to the bulk RR fields can be
read off from the Wess-Zumino
term~\cite{Kraus:2000nj,Jones:2002si,Sen:2003tm,Kennedy:1999nn}
and, for D$p{\bar {\rm D}}p$, it is possibly be extended as
\begin{eqnarray}
S_{\rm WZ}&=& \mu ~{\rm Str}\int_{\Sigma_3}C_{\rm RR}\wedge{\rm
exp}\left(\begin{array}{cc} F^{1}-T\bar{T} & i^{3/2}~DT \\
-i^{3/2}~\overline{DT} & F^{2}-\bar{T}T
\end{array}\right)
\nonumber \\
&=&\mu\int_{\Sigma_3}e^{-T\bar{T}}C_{\rm RR}\wedge(2C+iDT\wedge
\overline{DT})
\nonumber \\
&=&2\mu\int_{\Sigma_3}C_{\rm RR}\wedge dr\wedge d\theta
~\frac{d}{dr}\left[e^{-\tau^{2}}\left(C_\theta-\frac{n}{2}\right)\right]
, \label{rtwz}
\end{eqnarray}
where $\mu$ is a real constant and the supertrace ${\rm Str}$ is
defined to be a trace with insertion of $\sigma_3$. Note that
$C\equiv C_{\mu\nu}dx^\mu\wedge dx^\nu$ and
$C_{\mu\nu}=(F^1_{\mu\nu}-F^2_{\mu\nu})/2$. For the D2${\bar {\rm
D}}$2 system, it is obvious that all the singular and regular,
global ($C_\theta =0$) and local D-vortices on the D2${\bar {\rm
D}}$2 carry only D0 charge. In addition,
%
%
the total D0 charge ($=4\pi\mu\int dr
\frac{d}{dr}[e^{-\tau^{2}}\left(C_\theta-\frac{n}{2}\right)]=2\pi
\mu n $) is exactly proportional to the vorticity $n$ (or the
quantized magnetic flux (\ref{flu}) for the local vortex)
irrespective of the nature of the D-vortices, global or local, as
expected.
%

An extension from D-vortices to straight D-strings is
straightforwardly made by considering the coincidence limit of
D3${\bar {\rm D}}$3-system (with fundamental strings) instead of
the D2${\bar {\rm D}}$2. Since our analysis is based on the
DBI-type effective field theory action (\ref{dea}), the matrices
$X_{\mu\nu}$ (\ref{Xm}) and $X_{\mu\nu}^{\pm}$ (\ref{Xpm}) inside
the actions (\ref{acg}) and (\ref{acg1}) become $4\times
4$-matrices with coordinate $(t,r,\theta,z)$ instead of $3\times
3$ matrices. For straight D-strings stretched along the
$z$-direction, additional components of the matrices can be
assumed to have $X_{tz}=X_{tz}^{\pm}=0$ and
$X_{zz}=X_{zz}^{\pm}=1$ so that $S/\int dz$ is the same as the
actions (\ref{acg}) and (\ref{acg1}) due to translational symmetry
along the $z$-direction. Then the one-dimensional RR-charge is
derived, and the stringy objects are identified with singular and
regular D1-branes (D-strings).

In this paper we obtained singular and regular D-strings
(D1-branes) given by global and local vortex-string solutions.
Another attractive stringy configuration is DF-string  or
$(p,q)$-string (composite of D1F1), which also plays an important
role as a cosmic string~\cite{Copeland:2003bj}. In the scheme of
effective field theory, the simplest straight DF-string can be
achieved by adding another component of Born-Infeld type U(1)
gauge field, i.e., it is localized $E_{z}(r)=F_{tz}(r)$. If we
find a string solution with the RR-charge and the fundamental
string charge density $\Pi_{z}$ confined along the string, the
obtained stringy objects must be straight DF-string.

If we turn on angular component of the electric field $E_{\theta}=F_{0\theta}$,
the classical equations of motion force it to be constant
and are not likely to
support such static vortex solutions. In terms of string theory,
this impossibility seems natural since existence of such constant $E_{\theta}$
implies distribution of closed strings melted on the D-brane to infinity
with constant density.

In curved spacetime the obtained D-strings become candidates of cosmic
superstrings, i.e., they are straight cosmic D-strings.
The spacetime structure formed by these D-strings is of interest
in cosmology, and the first specific question is emergence of conical
space at asymptotic region when the cosmological constant vanishes.
Cosmic D-strings are mostly generated after the brane and antibrane meet,
which means post inflationary era since the inflation on brane-antibrane
occurs before the brane and antibrane meet.

\setcounter{equation}{0}
\section{Conclusion}
In this paper we considered DBI-type effective action of a complex
tachyon and gauge fields of U(1)$\times$U(1) symmetry, describing
brane-antibrane system with fundamental strings. In the
coincidence limit of D2${\bar {\rm D}}$2, static vortex solutions
are obtained. Without DBI electromagnetic field, there exist only
singular static global and local D-vortex solutions. When the
radial component of electric field is turned on, we found regular
static global and local D-vortex solutions. The obtained
point-like D-vortex configurations are naturally embedded in
straight stringy solutions in D3${\bar {\rm D}}$3-system, and are
identified with D-strings (D1-branes). If the obtained macroscopic
D-strings are gravitating, they become naturally candidates of
cosmic D-strings in the early Universe.

We conclude the paper with a few discussions for further study.
First, we only considered D${\bar {\rm D}}$ in the coincidence limit
through this paper, however it is intriguing to take into account
the effect coming from separation of D and ${\bar {\rm D}}$. In the effective
field theory, it means inclusion of the transverse coordinates and dynamics
of cosmological time evolution, e.g., inflation.
Second, in the superstring theory setting, the worldvolume of
D3${\bar {\rm D}}3$-system comprises a noncompact (1+3)-dimensional bulk
in ten dimensions. The remaining six-dimensions are assumed to be
compactified and their size effect was completely neglected in this work.
Since the tension of D-branes is naturally in string scale,
a consistent approach including KK-modes makes our analysis more concrete.
Third, Nielsen-Olesen vortices of Abelian-Higgs model are scattered to 90
degrees and this property leads to intercommutation (reconnection)
of two colliding cosmic strings.
But, according to Ref.~\cite{Copeland:2003bj,Jackson:2004zg},
colliding DF-strings may result in a connected tree structure
composed of a pair of trilinear vertices
and it can finally form a
cosmic DF-string network. This issue was dealt in the context of Yang-Mills
theory~\cite{Hashimoto:2003pu}
but it needs further study by using the obtained D(F)-strings
and the DBI-type nonlocal effective action.

\section*{Acknowledgments}
We would like to thank Jihoon Lee, A. Sen, and H. Tye for helpful discussions.
This work is the result of research activities (Astrophysical Research
Center for the Structure and Evolution of the Cosmos (ARCSEC))
supported by Korea Science $\&$ Engineering Foundation (Y.K.).

\end{document}